\documentclass[prd,12pt,nofootinbib,twoside,raggedbottom]{revtex4-1}
\usepackage{amsmath}
\usepackage{graphicx}
\usepackage{float}
\usepackage{color}
\usepackage{ulem}
\usepackage{hyperref}
\usepackage{array}
\usepackage{multirow}
\allowdisplaybreaks

\newcommand{\eeww}{e^+e^-\to W^+W^-}
\newcommand{\rac}{{\scshape RacoonWW}}
\newcommand{\yfs}{{\scshape YFSWW3}}
\newcommand{\Obs}[2]{\widehat{O}_{#1}^{\text{#2}}}
\newcommand{\db}{\bar\delta}

\newcommand{\delb}[1]{\bar\delta^{\text{#1}}}
\newcommand{\ALR}{A_{\text{LR}}}
\newcommand{\AbLR}{\overline{A}_{\text{LR}}}

\newcommand{\almz}{\alpha(m_Z)}
\newcommand{\alsmz}{\alpha_s(m_Z)}
\newcommand{\sineff}[1]{\sin^2\theta_{\text{eff}}^{#1}}
\newcommand{\sw}{s_\theta}
\newcommand{\cw}{c_\theta}

\newcommand{\be}{\begin{equation}}
\newcommand{\bea}{\begin{eqnarray}}
\newcommand{\beq}[1]{\begin{equation}\label{#1}}
\newcommand{\beqa}[1]{\begin{eqnarray}\label{#1}}
\newcommand{\eq}[1]{Eq.~(\ref{#1})}
\newcommand{\eqs}[2]{Eqs.~(\ref{#1}) and~(\ref{#2})}

\newcommand{\ee}{\end{equation}}
\newcommand{\eea}{\end{eqnarray}}
\newcommand{\eeq}{\end{equation}}
\newcommand{\eeqa}{\end{eqnarray}}

\newcommand{\diff}{\mathrm{d}}

\newcommand{\Lag}{\mathcal{L}}
\newcommand{\Op}[1]{\mathcal{O}_{#1}}
\newcommand{\Ord}[1]{\mathcal{O}\left(#1\right)}

\begin{document}

\begin{flushright}
\begin{small}
MCTP-15-14
\end{small}
\end{flushright}

\title{Status and prospects of precision analyses with $e^+e^-\to W^+W^-$}
\author{James~D.~Wells and Zhengkang~Zhang}
\affiliation{Michigan Center for Theoretical Physics, Department of Physics, University of Michigan, Ann Arbor, MI 48109, USA}
\date{\today}

\begin{abstract}
$e^+e^-\to W^+W^-$ is an integral part of the global precision analysis program which is becoming more relevant after the discovery of the Higgs boson. We analyze the current situation of precision calculations of inclusive $e^+e^-\to W^+W^-$ observables, and study the prospects of incorporating them into the framework of global precision electroweak analyses in light of per-mil-level cross section measurements at proposed future facilities. We present expansion formulas for the observables, making the dependence on the inputs clear. Also, the calculation of new physics effects is demonstrated in the effective field theory framework for universal theories. We go beyond the triple-gauge-couplings parametrization, and illustrate the complementarity of $e^+e^-\to W^+W^-$ and other precision data at the observables level.
\end{abstract}
\maketitle
\tableofcontents
\thispagestyle{empty}

\newpage
\section{Introduction}

The recent discovery of the Higgs boson~\cite{Aad:2012tfa,Chatrchyan:2012ufa} has initiated intense interest in precision studies of its properties. There have been various attempts to connect Higgs phenomenology to previous precision electroweak measurements, in order to see what windows for new physics are still open and can be probed by precision Higgs measurements~\cite{Masso:2012eq,Corbett:2012ja,Falkowski:2013dza,Contino:2013kra,Corbett:2013pja,Grinstein:2013vsa,Ciuchini:2013pca,Elias-Miro:2013mua,Pomarol:2013zra,Elias-Miro:2013eta,Trott:2014dma}. There are two facts behind this trend. First, deviations from the Standard Model (SM) in the Higgs and electroweak sectors are correlated if the recently discovered Higgs boson is part of an $SU(2)_L$ doublet. To fully exploit the power of precision analyses we should take advantage of all the strong constraints on new physics from precision electroweak studies. Second, it has been realized that in the effective field theory (EFT) framework, which is the only consistent model-independent approach to new physics with minimal assumptions~\cite{Degrande:2012wf,Willenbrock:2014bja,Henning:2014wua}, it is important to use a complete operator basis~\cite{Grzadkowski:2010es,Buchalla:2012qq,Buchalla:2013rka,Elias-Miro:2013mua,Elias-Miro:2013eta}. Therefore, previous EFT calculations of precision electroweak observables should be recast into a uniform framework for a global analysis. In this context, a crucial role is played by $\eeww$. Despite the lower experimental precision compared with the most precisely measured observables, $\eeww$ offers a unique window to test the SM and probe new physics effects due to its sensitivity to triple-gauge couplings (TGCs) which are difficult to access otherwise.

On the other hand, precision studies of $W$-pair production are important in their own right. Such studies have been carried out for $e^+e^-$ collisions at LEP2~\cite{Beenakker:1996kt,Bardin:1997gc,Grunewald:2000ju,Schael:2013ita}, for $p\bar p$ collisions at the Tevatron~\cite{Abazov:2012ze}, and more recently for $pp$ collisions at the LHC~\cite{ATLAS:2012mec,Chatrchyan:2013yaa}. The clean experimental environment at LEP2 allowed $\eeww$ cross sections to be measured at the $\sim1\%$ level up to $\sqrt{s}=207$ GeV, and agreements were found with SM predictions with a similar precision~\cite{Denner:1997ia,Denner:1999gp,Denner:1999kn,Denner:1999dt,Denner:2000bj,Denner:2001zp,Denner:2002cg,Jadach:1995sp,Jadach:1996hi,Jadach:1998tz,Jadach:2000kw,Jadach:2001uu}. Historically, analyses have been done in the language of TGC parameters~\cite{Hagiwara:1986vm}. In this context, LEP2 data still provide the most stringent constraints on anomalous TGCs, with LHC measurements just starting to become competitive\footnote{The $W\gamma$ and $WZ$ channels are also studied at the LHC to put limits on anomalous TGCs~\cite{Aad:2012twa,Chatrchyan:2012bd,Aad:2013izg,Chatrchyan:2013fya,Aad:2014mda}.}. Future high-luminosity $e^+e^-$ colliders, operating at center-of-mass energies from $W^+W^-$ threshold up to 1~TeV or even beyond, will enable per-mil-level cross section measurements in a wide range of $\sqrt{s}$ and push the precision frontier on $\eeww$ studies much further~\cite{Baer:2013cma,Baak:2013fwa,Moortgat-Picka:2015yla,Fujii:2015jha,Barklow:2015tja}. Accordingly, progress has been made after LEP2 on precision calculations toward the per-mil precision goal~\cite{Denner:2005es,Denner:2005fg}. The promising experimental progress calls for reassessment of the role of $e^+e^-\to W^+W^-$ in the precision program, both as a consistency test of the SM and an indirect probe of new physics.

In this paper, we revisit the calculation of several $\eeww$ observables, both in the SM and in the presence of new physics, in the expansion framework of Ref.~\cite{Wells:2014pga}. From the point of view of testing the SM, at present our best knowledge of the compatibility of the electroweak SM with data comes from global analyses of $Z$-pole observables and $m_W$, all of which have been very well-measured~\cite{ALEPH:2005ab,Group:2012gb} and precisely-calculated~\cite{Arbuzov:2005ma,Baak:2014ora}; see e.g.\ Refs.~\cite{Eberhardt:2012gv,Ciuchini:2013pca,Baak:2014ora,Ciuchini:2014dea,Ellis:2014jta,Agashe:2014kda} for recent global fits. It is interesting to ask whether improved measurements of $\eeww$ will play a complementary role in such global analyses (aside from a better determination of $m_W$). To answer this question we need to gather information on how the $\eeww$ observables depend on the inputs of the calculations, which has not received much attention in the past. The expansion formulas we present will make the answer transparent.

To demonstrate the role of $\eeww$ in probing new physics, we adopt the EFT approach. We will focus on an important class of new physics scenarios, the so-called ``universal theories''~\cite{Barbieri:2004qk}, for illustration. The results will be presented in a way that allows other precision constraints to be easily incorporated. Many previous studies used the reported experimental values for the TGCs to constrain the EFT parameter space, but there is a caveat related to the assumptions made when extracting the TGCs~\cite{Trott:2014dma}. In this regard, Ref.~\cite{Trott:2014dma} refers to the TGCs as ``constructed observables'', and points out that extreme care must be taken when relating constructed observables to EFT parameters. Here we take a different approach by working with well-defined {\it physical} observables which are free from such subtleties. A similar analysis has been done in Ref.~\cite{Buchalla:2013wpa} without discussing the interplay with other precision measurements.

We will begin in Sec.~\ref{sec:obs} by defining several $\eeww$ observables and reviewing their calculations in the SM. Sections~\ref{sec:param} and~\ref{sec:universal} are devoted to the SM and new physics aspects of precision analyses mentioned above, respectively. In Sec.~\ref{sec:conclusions} we conclude.

\section{Observables and Standard Model calculations}\label{sec:obs}

At leading order (LO), the process $\eeww$ is calculated in the SM from the $s$-channel $Z/\gamma$ exchange and $t$-channel neutrino exchange diagrams, known as the CC03 diagrams~\cite{Bardin:1997gc,Grunewald:2000ju}. Some LO results that will be used later are collected in Appendix~\ref{app:LO}. However, the notion of ``$e^+e^-\to W^+W^-$ observables'' is not well-defined once we go beyond LO to include finite $W^\pm$ width effects and radiative corrections. Strictly speaking, what we refer to as ``$e^+e^-\to W^+W^-$ observables'' should be understood as a shorthand for ``$e^+e^-\to 4f$ observables with $4f$ compatible with intermediate $W^+W^-$''. Even at tree level, there are diagrams involving only single intermediate $W^+$ or $W^-$ that contribute to $e^+e^-\to 4f$. They are known as CC11 diagrams that are not in the CC03 class~\cite{Bardin:1997gc,Grunewald:2000ju}, and interfere with the CC03 diagrams with intermediate $W^+W^-$. We will include all the CC11 diagrams in the calculation of SM predictions in Sec.~\ref{sec:param}, as opposed to the calculations adopted by LEP2 analyses where only CC03 diagrams are included~\cite{Schael:2013ita}. One should keep in mind that with finite $W^\pm$ width, only the sum of the complete set of CC11 diagrams is gauge invariant.

The $\eeww$ observables we consider are the polarized total cross sections $\sigma_L$, $\sigma_R$, defined as the cross sections with left- or right-handed incoming electron and unpolarized incoming positron. From these, the unpolarized total cross section
\be
\sigma = \frac{1}{2}(\sigma_L+\sigma_R),
\ee
and the left-right asymmetries
\be
\ALR\equiv \frac{\sigma_L-\sigma_R}{\sigma_L+\sigma_R} = 1-\frac{\sigma_R}{\sigma},\quad
\AbLR\equiv 1-\ALR = \frac{\sigma_R}{\sigma}
\ee
can be derived. $\AbLR$ may be more convenient than $\ALR$ because the latter is very close to 1. We will focus on these inclusive observables without applying kinematic cuts\footnote{The numerical difference between our reference values for $\sigma$ in Sec.~\ref{sec:param} and the results in Refs.~\cite{Denner:2002cg,Denner:2005es} is mostly due to the separation cuts applied in the latter.}. Other observables can be extracted from differential distributions, which can presumably be distorted by new physics effects. For example, the forward-backward asymmetry defined for the outgoing $W^-$ is often considered; see e.g.~\cite{Buchalla:2013wpa} for an EFT study of differential cross sections at LO. However, once nonfactorizable radiative corrections and off-shell effects are taken into account, experimental subtleties arise related to the kinematic reconstruction of $W^\pm$ from the $4f$ final states, which should be carefully studied and is beyond the scope of the present paper.

The state-of-the-art calculations of $e^+e^-\to W^+W^-$ cross sections in the LEP2 era incorporated $\Ord{\alpha}$ radiative corrections in the double-pole approximation (DPA), and were implemented in dedicated programs \rac~\cite{Denner:1997ia,Denner:1999gp,Denner:1999kn,Denner:1999dt,Denner:2000bj,Denner:2001zp,Denner:2002cg} and \yfs~\cite{Jadach:1995sp,Jadach:1996hi,Jadach:1998tz,Jadach:2000kw,Jadach:2001uu}. Later the complete $\Ord{\alpha}$ radiative corrections were calculated for the four-fermion final states $\nu_\tau \tau^+ \mu^- \bar\nu_\mu$, $u \bar d \mu^- \bar\nu_\mu$ and $u \bar d s \bar c$~\cite{Denner:2005es,Denner:2005fg}. However, the latter calculation is not yet available as public codes. So for the purpose of illustration we will present the results as calculated within the DPA implemented in the program \rac~\cite{Denner:2002cg}. The unpolarized total cross section in the DPA agrees with the complete $\Ord{\alpha}$ result within 0.3\% for $\sqrt{s}$ from 200~GeV to 500~GeV~\cite{Denner:2005es}. To achieve better precision suitable for studies at future colliders, the results presented here are expected to be updated once more up-to-date codes become available. We also remark that to achieve better-than-per-mil accuracy, even the complete $\Ord{\alpha}$ calculation needs to be supplemented by higher-order Coulomb corrections near the threshold, higher-order Sudakov logarithms for $\sqrt{s}\gtrsim500$~GeV, and QCD effects~\cite{Denner:2005es,Kuhn:2007ca,Baak:2013fwa}.

We will consider two benchmark center-of-mass energies $\sqrt{s}=200~\text{GeV}$ and $500~\text{GeV}$, where the DPA works reasonably well. Among all the four-fermion final states compatible with intermediate $W^+W^-$, we will focus on $u \bar d \mu^- \bar\nu_\mu$ for illustration\footnote{Theory calculations are conventionally formulated in terms of partonic final states, though experimentally jets instead of quarks are observed. In this sense what we call ``observables'' are not yet experimentally observed quantities, but are directly related to the latter when we sum inclusive quark contributions that form jets.}. This channel is representative of the mixed leptonic and hadronic decay channels from $W^+W^-$, which are expected to have high selection efficiency and a low background. Separate calculations may be needed if one is interested in other channels due to additional diagrammatic contributions and the necessary inclusion of finite-electron-mass effects in the case of final-state $e^\pm$.

\section{Parametric dependence and uncertainties}\label{sec:param}

We will present the results of the SM calculations in the form of expansion formulas as in Ref.~\cite{Wells:2014pga}, which make clear the parametric dependence and uncertainties. To briefly review the formalism, we note that the SM predicts each observable $\Obs{i}{}$ as a function of the Lagrangian parameters. A more convenient way to arrange the calculation, which is commonly adopted in precision electroweak analyses, is to eliminate the input Lagrangian parameters in favor of the same number of very well-measured observables $\{\Obs{i'}{}\}$ (as in Ref.~\cite{Wells:2014pga}, we use primed indices for input observables and unprimed indices for output observables). Then the analysis involves only physical observables, so that all results have unambiguous meanings, and comparison with experiment is straightforward. The calculation is further simplified by expanding the theory (SM) predictions about some reference point and keeping terms up to linear order:
\beq{OSM}
\Obs{i}{th} = \Obs{i}{SM} = \Obs{i}{ref} + \sum_{i'} \left.\frac{\partial \Obs{i}{SM}}{\partial \Obs{i'}{}}\right|_{\Obs{i'}{}=\Obs{i'}{ref}} \Bigl(\Obs{i'}{}-\Obs{i'}{ref}\Bigr)
= \Obs{i}{ref} \biggl(1+\sum_{i'} c_{i,i'}\delb{SM}\Obs{i'}{}\biggr),
\eeq
where
\beq{delbSM}
\delb{SM}\Obs{i'}{} \equiv \frac{\Obs{i'}{}-\Obs{i'}{ref}}{\Obs{i'}{ref}}
\eeq
is the fractional shift of the input observable $\Obs{i'}{}$ with respect to its reference value, and
\be
c_{i,i'} \equiv \frac{\Obs{i'}{ref}}{\Obs{i}{ref}} \left.\frac{\partial \Obs{i}{SM}}{\partial \Obs{i'}{}}\right|_{\Obs{i'}{}=\Obs{i'}{ref}}
\ee
represents the resulting fractional shift of the output observable $\Obs{i}{}$ calculated in the SM. The superscript ``SM'' in \eq{delbSM} indicates that the shift can be associated with adjusting the SM Lagrangian parameters, which should be distinguished from corrections due to new physics; cf.~\eq{npshift}. These expansion coefficients characterize the parametric dependence of the calculated observables on the input observables, as long as the expansion up to first order is adequate. This is the case for most practical purposes now that the mass of the Higgs boson is known to subpercent level, and so all $|\delb{SM}\Obs{i'}{}|$ have to be much smaller than unity.

The standard set of input observables commonly adopted in precision electroweak analyses consists of the masses of the $Z$ boson, the top quark, and the Higgs boson $m_Z, m_t, m_H$, the Fermi constant $G_F$, and the couplings $\almz,\alsmz$. In the case of $\eeww$, however, it is more convenient to use the $W$ boson mass $m_W$ in place of $\almz$ as a calculational input. Thus, we first extract the expansion coefficients with respect to the input observables
\be
\{\Obs{i'}{}\} = \{ m_Z, G_F, m_W, m_t, \alsmz, m_H \},
\ee
and then transform the results into the standard basis
\be
\{\Obs{i'}{}\} = \{ m_Z, G_F, \almz, m_t, \alsmz, m_H \}.
\ee
The SM predictions for the observables take the following forms in the two basis:
\bea
\Obs{i}{SM} &=& \Obs{i}{ref} \Bigl[1 + c_{i,m_Z} \delb{SM}m_Z + c_{i,G_F} \delb{SM}G_F + c_{i,m_W} \delb{SM}m_W \nonumber\\
&&\qquad\quad + c_{i,m_t} \delb{SM}m_t + c_{i,\alpha_s} \delb{SM}\alsmz + c_{i,m_H} \delb{SM}m_H \Bigr]\label{expc}\\
&=& \Obs{i}{ref} \Bigl[1 + d_{i,m_Z} \delb{SM}m_Z + d_{i,G_F} \delb{SM}G_F + d_{i,\alpha} \delb{SM}\almz \nonumber\\
&&\qquad\quad + d_{i,m_t} \delb{SM}m_t + d_{i,\alpha_s} \delb{SM}\alsmz + d_{i,m_H} \delb{SM}m_H \Bigr].\label{expd}
\eea
The transformation from the $c_{i,i'}$ coefficients to the $d_{i,i'}$ coefficients
\bea
d_{i,\alpha} &=& c_{i,m_W} d_{m_W,\alpha},\\
d_{i,i'} &=& c_{i,i'} + c_{i,m_W} d_{m_W,i'} \,\,\text{for}\,\, \Obs{i'}{}\in\{m_Z, G_F, m_t, \alsmz, m_H\}
\eea
can be derived similarly as in Section~3.4 of Ref.~\cite{Wells:2014pga}.

We adopt the following reference values for the input observables:
\bea
m_Z^{\text{ref}} &=& 91.1876~\text{GeV},\\
G_F^{\text{ref}} &=& 1.1663787\times10^{-5}~\text{GeV}^{-2},\\
m_W^{\text{ref}} &=& 80.3614~\text{GeV},\\
m_t^{\text{ref}} &=& 174.17~\text{GeV},\\
\alsmz^{\text{ref}} &=& 0.1185,\\
m_H^{\text{ref}} &=& 125.9~\text{GeV}.
\eea
These, according to the formulas presented in~\cite{Wells:2014pga}, correspond to
\be
\almz^{\text{ref}} = 7.81861\times10^{-3} =1/127.900.
\ee
The final results of the SM predictions for the five $\eeww$ observables are the expansion formulas \eqs{expc}{expd}, with the reference values and expansion coefficients listed in Tables~\ref{table:c} and~\ref{table:d}, respectively, for the benchmark center-of-mass energies $\sqrt{s}=200~\text{GeV}, 500~\text{GeV}$. The error bars quoted contain (possibly overestimated) Monte Carlo errors only, while truncation errors from numerical differentiation are expected to be smaller. Further technical details of the calculation can be found in Appendix~\ref{sec:tech}.
\begin{table}
\begin{center}
  \begin{tabular}{|>{\footnotesize}c|>{\footnotesize}c|>{\footnotesize}c|>{\footnotesize}c>{\footnotesize}c>{\footnotesize}c>{\footnotesize}c>{\footnotesize}c>{\footnotesize}c|}
    \hline
    \hspace{1pt}$\sqrt{s}/$GeV\hspace{1pt} & \hspace{10pt}$\Obs{i}{}$\hspace{10pt} & \hspace{10pt}$\Obs{i}{ref}$\hspace{10pt} & \hspace{10pt}$c_{i,m_Z}$\hspace{10pt} & \hspace{10pt}$c_{i,G_F}$\hspace{10pt} & \hspace{10pt}$c_{i,m_W}$\hspace{10pt} & \hspace{10pt}$c_{i,m_t}$\hspace{10pt} & \hspace{10pt}$c_{i,\alpha_s}$\hspace{10pt} & \hspace{10pt}$c_{i,m_H}$\hspace{10pt} \\
    \hline
     & $\sigma_L$/fb & 1229.8(5) & 0.123(15) & 1.957(15) & 1.429(15) & 0.0038(29) & -0.0141(29) & 0.0006(29) \\
     & $\sigma_R$/fb & 13.593(5) & 17.48(8) & 1.981(13) & -17.03(8) & 0.1364(27) & -0.0138(27) & -0.041(5)  \\
    200 & $\sigma$/fb & 621.70(25) & 0.312(15) & 1.958(15) & 1.227(15) & 0.0052(29) & -0.0141(29) & 0.0002(29)  \\
     & $\ALR$ & 0.978136(12) & -0.3835(18) & -0.0005(4) & 0.4078(18) & -0.00293(9) & -0.00001(9) & 0.00092(14) \\
     & $\AbLR$ & 0.021864(12) & 17.17(8)  & 0.024(20) & -18.26(8) & 0.131(4) & 0.000(4) & -0.041(6) \\
    \hline
     & $\sigma_L$/fb & 553.48(22) & 0.341(18) & 2.022(18) & 2.936(19) & 0.0023(29) & -0.0120(29) & 0.0005(29) \\
     & $\sigma_R$/fb & 3.4687(13) & 14.93(7) & 2.098(12) & -11.04(7) & 0.1710(25) & -0.0104(25) & 0.0042(25) \\
    500 & $\sigma$/fb & 278.48(11) & 0.432(18) & 2.022(18) & 2.849(19) & 0.0034(28) & -0.0120(28) & 0.0005(28) \\
     & $\ALR$ & 0.987544(7) & -0.1826(10) & -0.00096(28) & 0.1751(10) & -0.00211(5) & -0.00002(5) & -0.00005(5) \\
     & $\AbLR$ & 0.012456(7) & 14.49(8) & 0.076(22) & -13.89(8) & 0.168(4) & 0.002(4) & 0.004(4) \\
    \hline
  \end{tabular}
\end{center}
\caption{Reference values and expansion coefficients for the $\eeww$ observables (for $u\bar d \mu^-\bar\nu_\mu$ final state) with respect to the input observables $\{m_Z, G_F, m_W, m_t, \alsmz, m_H\}$, to be used in \eq{expc}.}\label{table:c}
\end{table}
\begin{table}
\begin{center}
  \begin{tabular}{|>{\footnotesize}c|>{\footnotesize}c|>{\footnotesize}c|>{\footnotesize}c>{\footnotesize}c>{\footnotesize}c>{\footnotesize}c>{\footnotesize}c>{\footnotesize}c|}
    \hline
    \hspace{1pt}$\sqrt{s}/$GeV\hspace{1pt} & \hspace{10pt}$\Obs{i}{}$\hspace{10pt} & \hspace{10pt}$\Obs{i}{ref}$\hspace{10pt} & \hspace{10pt}$d_{i,m_Z}$\hspace{10pt} & \hspace{10pt}$d_{i,G_F}$\hspace{10pt} & \hspace{10pt}$d_{i,\alpha}$\hspace{10pt} & \hspace{10pt}$d_{i,m_t}$\hspace{10pt} & \hspace{10pt}$d_{i,\alpha_s}$\hspace{10pt} & \hspace{10pt}$d_{i,m_H}$\hspace{10pt} \\
    \hline
     & $\sigma_L$/fb & 1229.8(5) & 2.163(26) & 2.272(16) & -0.3077(33) & 0.0227(29) & -0.0154(29) & -0.0005(29) \\
     & $\sigma_R$/fb & 13.593(5) & -6.84(14)  & -1.767(22) & 3.668(17) & -0.0893(29) & 0.0026(27) & -0.028(5)  \\
    200 & $\sigma$/fb & 621.70(25) & 2.064(26) & 2.228(15) & -0.2643(32) & 0.0215(29) & -0.0152(29) & -0.0008(29)  \\
     & $\ALR$ & 0.978136(12) & 0.1988(32) & 0.0892(6) & -0.0878(4) & 0.00247(9) & -0.00040(9) & 0.00061(14) \\
     & $\AbLR$ & 0.021864(12) & -8.90(14)  & -3.995(27) & 3.933(17) & -0.111(4) & 0.018(4) & -0.027(6) \\
    \hline
     & $\sigma_L$/fb & 553.48(22) & 4.534(32) & 2.668(19) & -0.633(4) & 0.0412(29) & -0.0148(29) & -0.0017(29) \\
     & $\sigma_R$/fb & 3.4687(13) & -0.85(13) & -0.333(20) & 2.379(16) & 0.0246(27) & 0.0002(25) & 0.0127(25)  \\
    500 & $\sigma$/fb & 278.48(11) & 4.501(32) & 2.649(19) & -0.614(4) & 0.0411(28) & -0.0148(28) & -0.0017(28) \\
     & $\ALR$ & 0.987544(7) & 0.0674(17) & 0.03757(35) & -0.03771(21) & 0.00021(5) & -0.00019(5) & -0.00018(5) \\
     & $\AbLR$ & 0.012456(7) & -5.35(13) & -2.982(28) & 2.993(16) & -0.016(4) & 0.015(4) & 0.014(4) \\
    \hline
  \end{tabular}
\end{center}
\caption{Reference values and expansion coefficients for the $\eeww$ observables (for $u\bar d \mu^-\bar\nu_\mu$ final state) with respect to the input observables $\{m_Z, G_F, \almz, m_t, \alsmz, m_H\}$, to be used in \eq{expd}.}\label{table:d}
\end{table}

\begin{figure}
\centering
\includegraphics[width=3.2in]{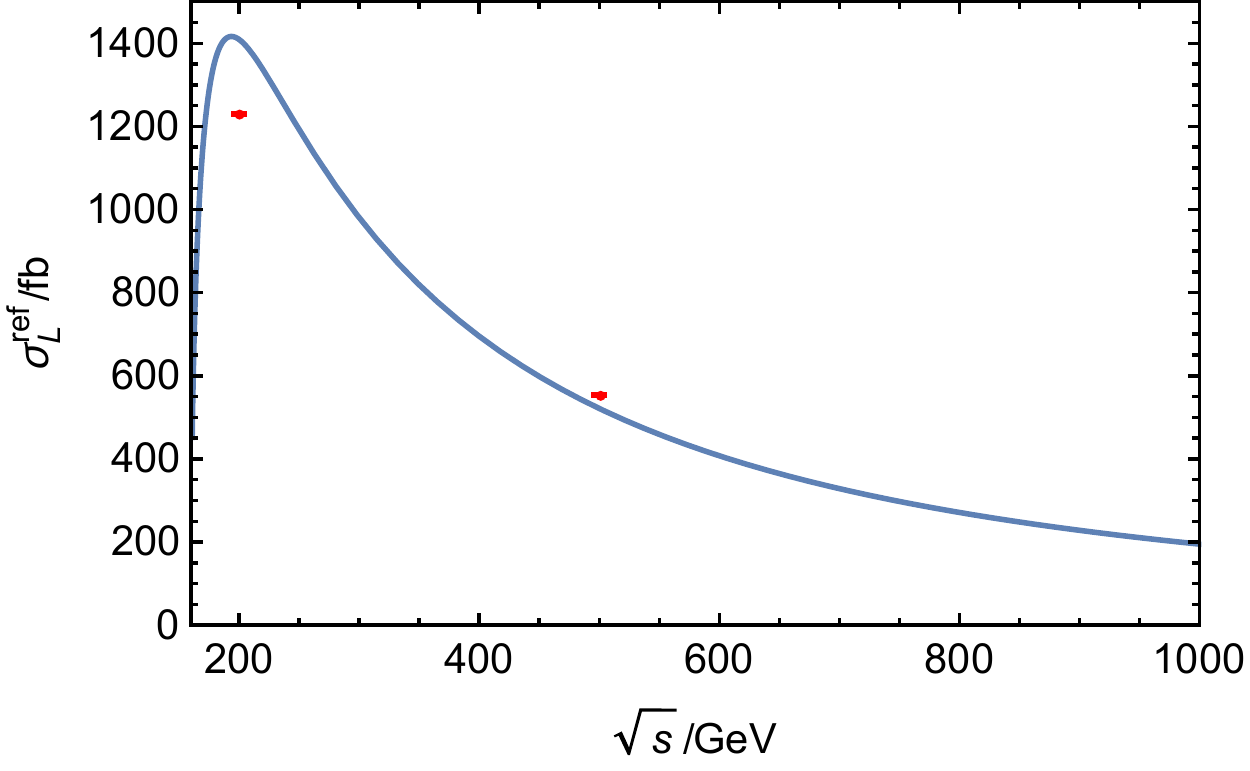}
\includegraphics[width=3.2in]{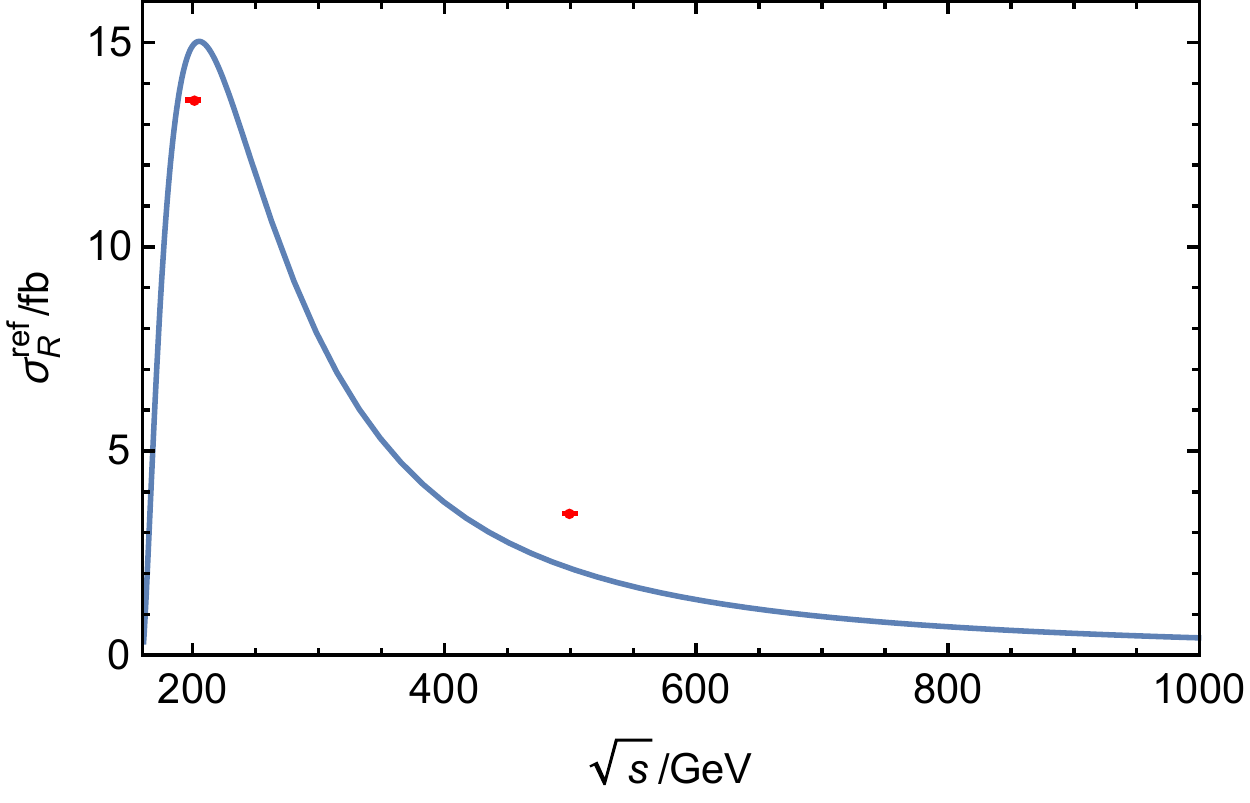}
\includegraphics[width=3.2in]{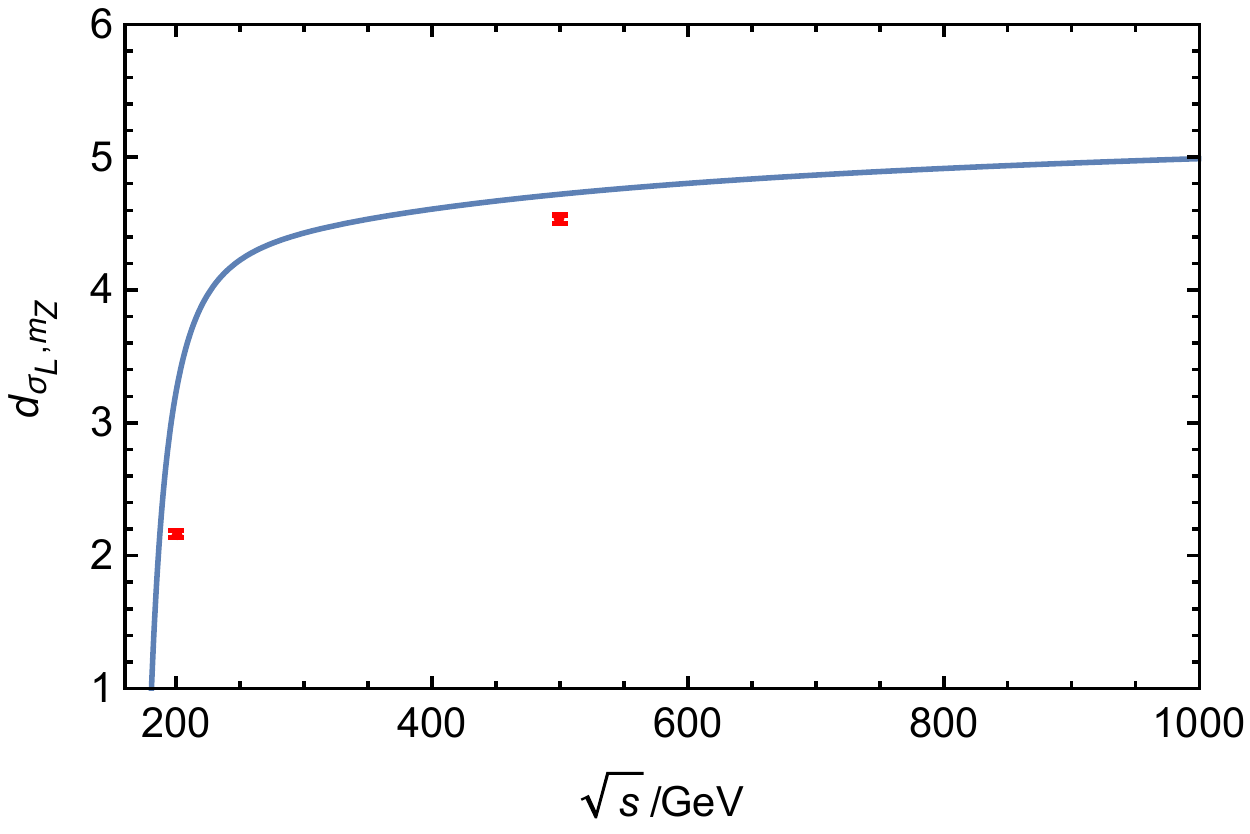}
\includegraphics[width=3.2in]{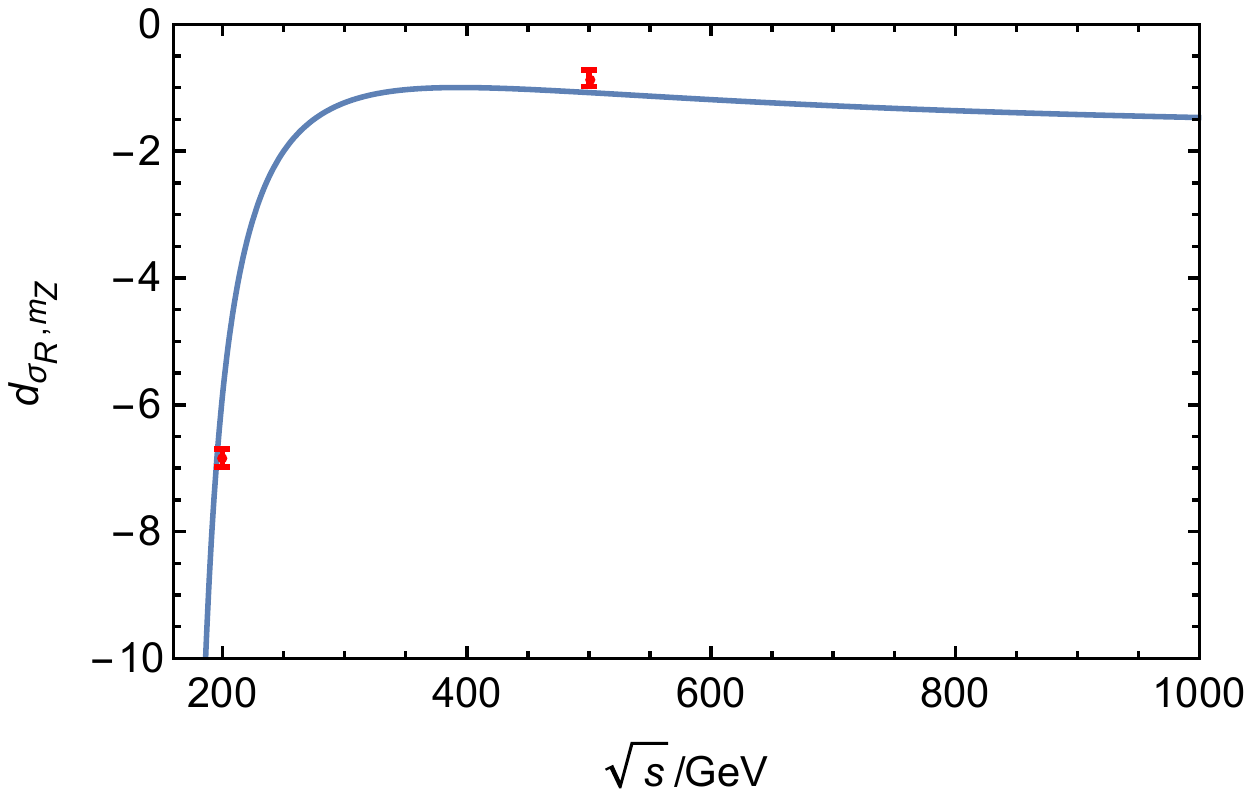}
\includegraphics[width=3.2in]{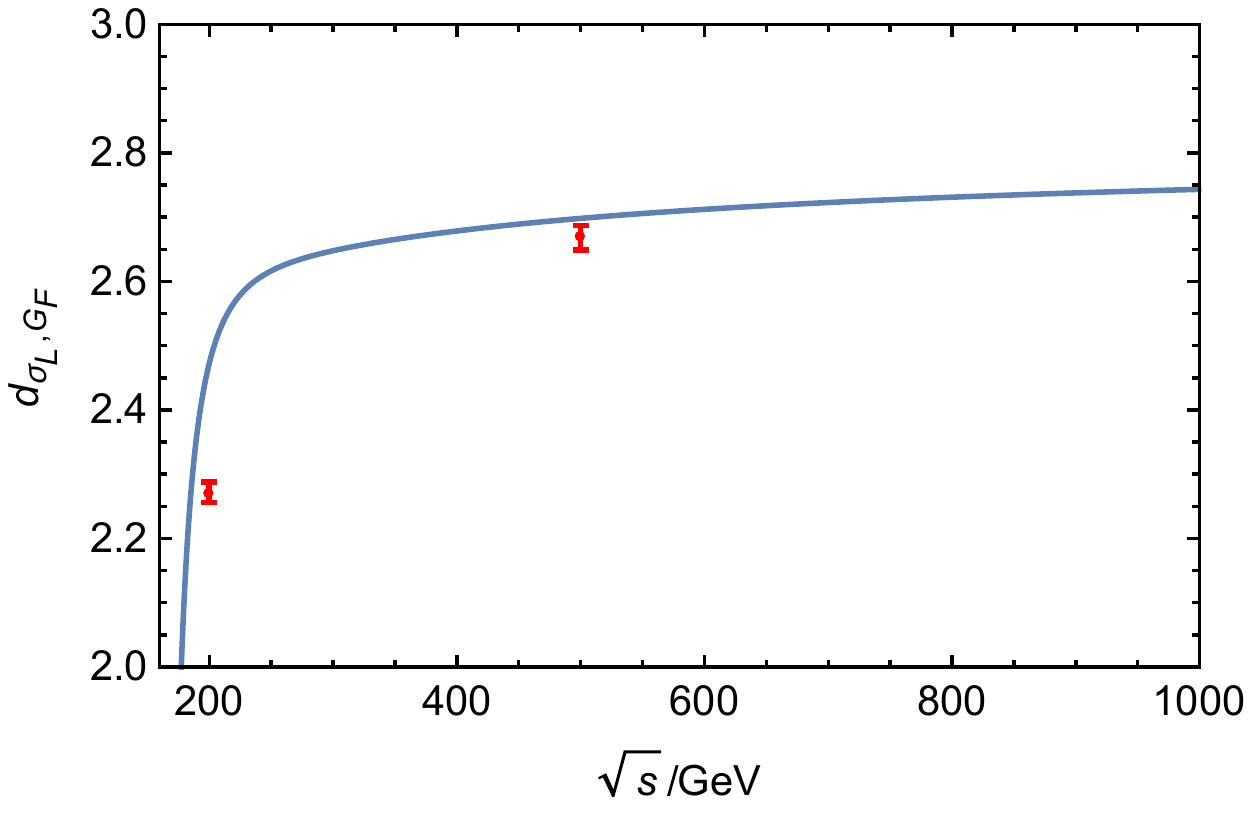}
\includegraphics[width=3.2in]{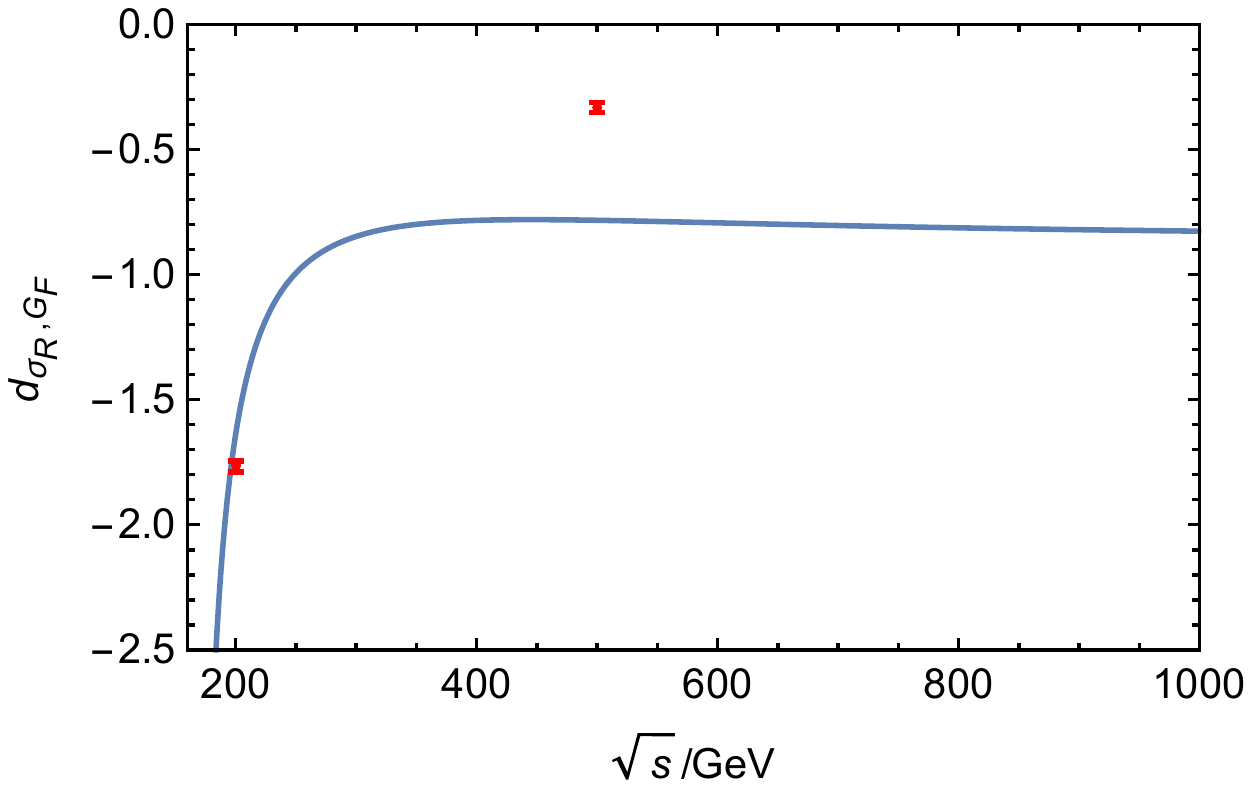}
\includegraphics[width=3.2in]{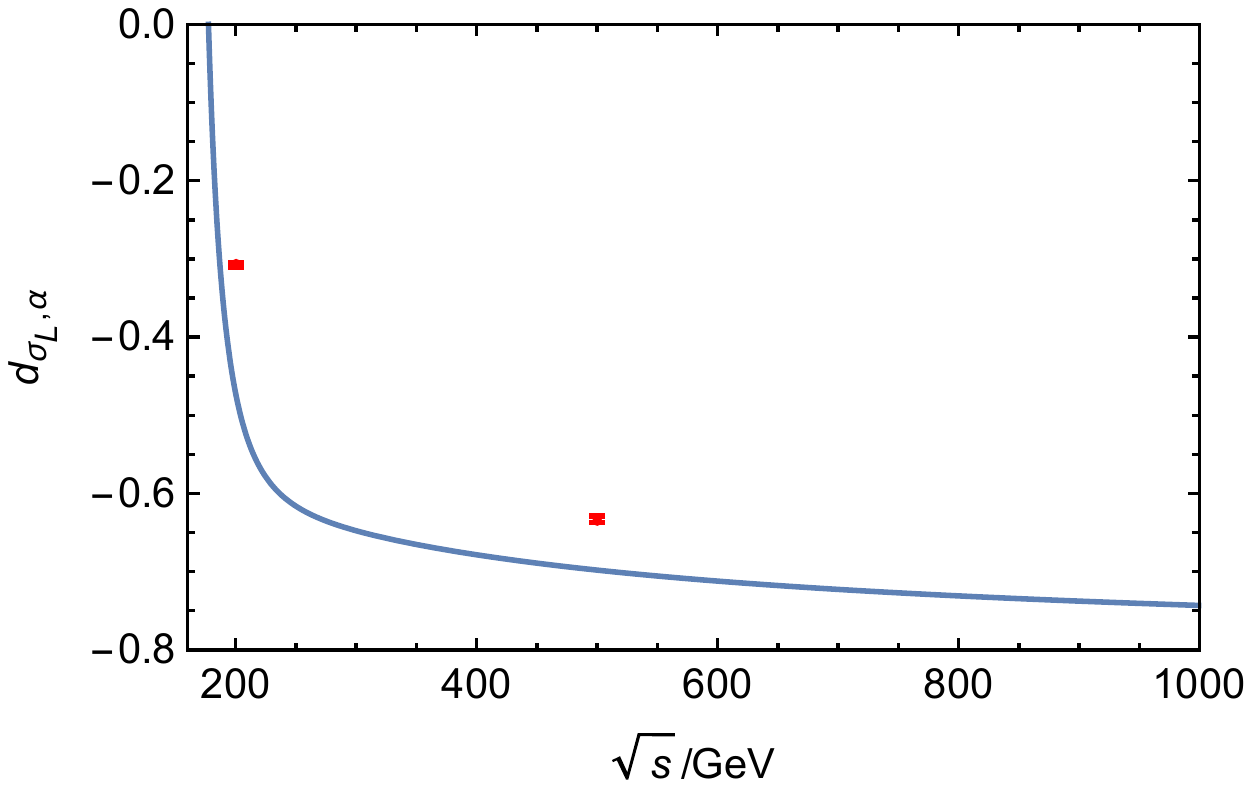}
\includegraphics[width=3.2in]{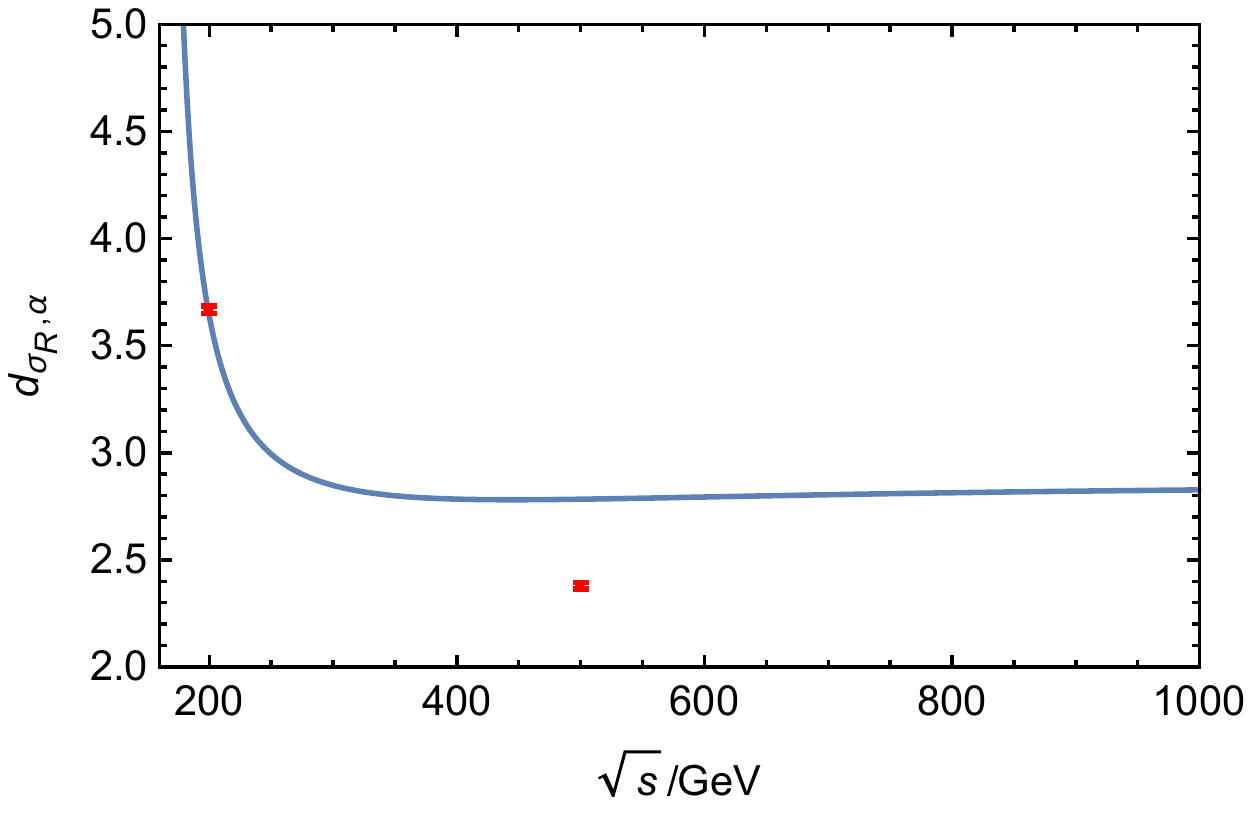}
\caption{Comparison of the \rac\ calculations (red data points) of $\sigma_{L,R}^\text{ref}$ and expansion coefficients with LO results (continuous curves) as functions of $\sqrt{s}$.}
\label{fig:NLOvsLO}
\end{figure}

To get an idea of the size of $\Ord{\alpha}$ radiative corrections, as well as the dependence on $\sqrt{s}$ beyond the two benchmark choices, we show in Fig.~\ref{fig:NLOvsLO} the comparison of the numerical results in Table~\ref{table:d} with LO results as functions of $\sqrt{s}$ for the two observables $\sigma_L$ and $\sigma_R$. The latter can be easily calculated analytically; see Appendix~\ref{app:LO}. Only $\Obs{i}{ref}$, $d_{i,m_Z}$, $d_{i,G_F}$, $d_{i,\alpha}$ are presented, since the other three expansion coefficients vanish at LO. Fig.~\ref{fig:NLOvsLO} shows that in most cases, the $\Ord{\alpha}$ corrections are at or below $\Ord{10\%}$ level. We note that while detailed discussions on the size of $\Ord{\alpha}$ corrections for $\Obs{i}{ref}$ can be found in the papers where these corrections are calculated, the plots for the expansion coefficients $d_{i,i'}$ are new. The latter provide complementary information because both $\Obs{i}{ref}$ and $d_{i,i'}$ are needed in precision analyses if one is not restricted to fixed values of the input observables. Also, $d_{i,i'}$ are essential in a consistent calculation of new physics effects, as we will see in the next section.

The results in Tables~\ref{table:c} and~\ref{table:d} reflect the accuracy implemented in \rac, namely up to $\Ord{\alpha}$ (with respect to LO) and within the DPA. They are expected to be updated in the future. However, even at present, these finite-accuracy expansion formulas are useful for the purpose of having a picture of parametric dependence and an estimate of parametric uncertainties, i.e.\ uncertainties from the input parameters (observables). With the experimental uncertainties of the input observables taken from Ref.~\cite{Wells:2014pga},
\bea
&&\Delta m_Z = 2.1\text{ MeV},\,\, \Delta G_F = 6\times10^{-12}\text{ GeV}^{-2},\,\, \Delta \almz = 8.6\times10^{-7},\nonumber\\
&&\Delta m_t = 0.87\text{ GeV},\,\, \Delta \alsmz = 0.0006,\,\, \Delta m_H = 0.4\text{ GeV},
\eea
the fractional parametric uncertainty in $\Obs{i}{}$ from input observable $\Obs{i'}{}$ is easily obtained by
\be
\frac{\Delta \Obs{i}{}}{\Obs{i}{}} = |d_{i,i'}| \frac{\Delta\Obs{i'}{}}{\Obs{i'}{ref}} \equiv \Delta_{i,i'}\cdot 10^{-4}.
\ee
The results for $\Delta_{i,i'}$ are listed in Table~\ref{table:parunc} for two representative observables $\sigma_R$ and $\sigma$.

\begin{table}
\begin{center}
  \begin{tabular}{|c|c|cccccc|}
    \hline
    \hspace{1pt}$\sqrt{s}/$GeV\hspace{1pt} & \hspace{10pt}$\Obs{i}{}$\hspace{10pt} & \hspace{10pt}$\Delta_{i,m_Z}$\hspace{10pt} & \hspace{10pt}$\Delta_{i,G_F}$\hspace{10pt} & \hspace{10pt}$\Delta_{i,\alpha}$\hspace{10pt} & \hspace{10pt}$\Delta_{i,m_t}$\hspace{10pt} & \hspace{10pt}$\Delta_{i,\alpha_s}$\hspace{10pt} & \hspace{10pt}$\Delta_{i,m_H}$\hspace{10pt} \\
    \hline
    \multirow{2}{1.5cm}{\centering 200} 
    & $\sigma_R$ & 1.57  & 0.01 & 4.07 & 4.46(15) & 0.13(14) & 0.89(17)  \\
     & $\sigma$ & 0.48 & 0.01 & 0.29 & 1.07(14) & 0.77(15) & 0.02(9)  \\
    \hline
    \multirow{2}{1.5cm}{\centering 500} 
    & $\sigma_R$ & 0.19 & 0.00 & 2.64 & 1.23(14) & 0.01(13) & 0.40(8) \\
     & $\sigma$ & 1.04 & 0.01 & 0.68 & 2.05(14) & 0.75(14) & 0.05(9) \\
    \hline
  \end{tabular}
\end{center}
\caption{Fractional parametric uncertainties from each input observable, in units of $10^{-4}$. For example, at $\sqrt{s}=200$~GeV, $\Delta\sigma/\sigma$  from $m_Z$ is $0.48\times10^{-4}$. These parametric uncertainties are negligible compared with experimental and theoretical uncertainties.}\label{table:parunc}
\end{table}

The parametric dependence and uncertainties for $e^+e^-\to W^+W^-$ observables are usually not discussed in the literature, but they provide important information if we put these observables into the broader context of precision electroweak analyses. In particular, a global $\chi^2$ fit is dominated by observables for which experimental and theoretical uncertainties are not much larger than parametric uncertainties, e.g.~the effective electroweak mixing angle $\sineff{\ell}$. In this regard, we find that the projected per-mil-level measurements and calculations of the $\eeww$ total cross section $\sigma$ are still not good enough compared with the very small parametric uncertainties. The latter are seen from Table~\ref{table:parunc} to be at the $10^{-4}$ level at present, and will be further reduced in the future with more precise measurements of the input observables. The parametric uncertainties are larger for $\sigma_R$, but are still expected to be much smaller than the experimental errors, given the limited statistics for $\sigma_R$ since $\sigma_R\ll\sigma$. We thus conclude that the $e^+e^-\to W^+W^-$ observables are likely to remain of peripheral importance in a global precision electroweak analysis as a compatibility test of the SM. Nevertheless, $\eeww$ can be a uniquely powerful probe of some new physics scenarios, as we will describe below. For this purpose, the results in this section indicate that new physics effects above the $\Ord{10^{-4}}$ level can be studied without letting the SM input observables float even in a global analysis.

\section{Probing universal theories with $e^+e^-\to W^+W^-$}\label{sec:universal}

\subsection{Universal theories and operator bases}

In the presence of new physics, an additional term $\xi_i$ is added to the right-hand-side of \eq{OSM}, which is defined as the fractional shift in $\Obs{i}{th}$ due to new physics calculated in terms of the {\it Lagrangian parameters}. It is assumed that $|\xi_i|\ll 1$ so that an expansion up to linear order is still adequate. To conform with the very precise measurements of the input observables, we adjust the SM Lagrangian parameters to keep
\be
\db\Obs{i'}{th} \equiv \frac{\Obs{i'}{th}-\Obs{i'}{ref}}{\Obs{i'}{ref}} = \delb{SM}\Obs{i'}{} + \xi_{i'}
\ee
small. Then the output observables are calculated as follows:
\beq{npshift}
\Obs{i}{th} = \Obs{i}{ref} \left(1+ \sum_{i'} d_{i,i'} \delb{SM}\Obs{i'}{} + \xi_i \right) = \Obs{i}{ref} \left(1+ \sum_{i'} d_{i,i'} \db\Obs{i'}{th} + \delb{NP}\Obs{i}{} \right),
\eeq
where
\be
\delb{NP}\Obs{i}{} \equiv \xi_i - \sum_{i'} d_{i,i'} \xi_{i'} = \xi_i - d_{i,m_Z}\xi_{m_Z} - d_{i,G_F}\xi_{G_F} - d_{i,\alpha}\xi_\alpha - \dots
\ee
In \eq{npshift}, the shift in $\Obs{i}{th}$ with respect to $\Obs{i}{ref}$ is consistently decomposed into the shift due to tuning the values of the input observables $\Obs{i'}{th}$ and the shift due to new physics. The latter is in turn decomposed into a direct contribution $\xi_i$ and an indirect contribution from undoing the shifts in the input observables.

We will apply this formalism to a popular class of new physics scenarios, the ``universal theories'', and investigate their effects on $\eeww$. Universal theories are defined by the assumption that new vector states, if there are any, couple to SM fermions only via the $SU(2)_L\times U(1)_Y$ currents~\cite{Barbieri:2004qk}. In other words, it is assumed that the only gauge interactions of the SM fermions, apart from QCD, have the form
\be
\Delta\Lag = g \bar W^a_\mu \Bigl(\bar l \gamma^\mu \frac{\sigma^a}{2} l + \bar q \gamma^\mu \frac{\sigma^a}{2} q\Bigr) + g' \bar B_\mu \sum_{f\in\{l,e,q,u,d\}} Y_f \bar f \gamma^\mu f,
\ee
where $\bar W^a_\mu$, $\bar B_\mu$ need not coincide with the SM gauge fields. Simple examples that qualify as universal theories include $W'$, $Z'$ models, where $\bar W^a_\mu$, $\bar B_\mu$ are mixtures of SM and new vector bosons. More complicated new physics models that are well-motivated (e.g.~little Higgs models and some extra-dimension models) can also be cast into this form~\cite{Barbieri:2004qk}.

If we further assume that the scale of new physics $\Lambda$ is somewhat higher than $\sqrt{s}$ (which is well-motivated for $\sqrt{s}\lesssim500$~GeV given the non-observation of particles beyond the SM so far, though the situation might change), their effects can be parametrized model-independently by an EFT,
\be
\Lag = \Lag_\text{SM} + \sum_i \frac{c_i}{v^2} \Op{i} + \Ord{\frac{v^4}{\Lambda^4}}\quad\text{where } c_i\sim\Ord{\frac{v^2}{\Lambda^2}}.
\ee
By organizing the EFT as an expansion in $v^2/\Lambda^2$, we have assumed the recently-discovered Higgs boson is part of an $SU(2)_L$ doublet. Relaxing this assumption leads to the nonlinear version of the EFT, which is more appropriately organized as a loop expansion~\cite{Buchalla:2013eza,Buchalla:2014eca,Buchalla:2012qq,Buchalla:2013rka,Contino:2013kra}. We will calculate $\delb{NP}\sigma_{L,R}$ at LO, namely $\Ord{\frac{v^2}{\Lambda^2}}$ terms arising from the interference between tree-level diagrams with one dimension-six operator insertion and LO diagrams in the SM. In particular, this means CP-odd operators do not enter. Also, in the limit $\frac{\Gamma_W}{m_W}\to0$, universal theories can modify $W^+W^-$ production but not $W^+W^-$ decay. Corrections to the LO results include $\Ord{\frac{v^4}{\Lambda^4}}$, $\Ord{\frac{v^2}{\Lambda^2}\frac{\Gamma_W}{m_W}}$, and $\Ord{\frac{v^2}{\Lambda^2}\alpha}$. Fig.~\ref{fig:NLOvsLO} can be viewed as indicating the small size of part of the $\Ord{\frac{v^2}{\Lambda^2}\alpha}$ corrections. Light fermion Yukawa couplings will be neglected throughout.

The defining assumption of universal theories translates into constraints imposed on the EFT. With proper field redefinitions, it is possible to transfer all the new physics effects into the bosonic sector. In other words, the effective operators generated by new physics involve SM boson fields only. Among all the bosonic operators respecting SM symmetries one can write down, the following ones contribute to $\eeww$ at LO:
\bea
\Op{T} &=& \frac{1}{2} (H^\dagger \overleftrightarrow{D}_\mu H)^2,\\
\Op{WB} &=& g' g H^\dagger \sigma^a H W^a_{\mu\nu} B^{\mu\nu},\\
\Op{2W} &=& -\frac{1}{2} (D^\mu W^a_{\mu\nu})^2,\\
\Op{2B} &=& -\frac{1}{2} (\partial^\mu B_{\mu\nu})^2,\\
\Op{W} &=& \frac{ig}{2} (H^\dagger \sigma^a \overleftrightarrow{D}^\mu H) D^\nu W^a_{\mu\nu},\\
\Op{B} &=& \frac{ig'}{2} (H^\dagger \overleftrightarrow{D}^\mu H) \partial^\nu B_{\mu\nu},\\
\Op{HW} &=& ig (D^\mu H)^\dagger \sigma^a (D^\nu H) W^a_{\mu\nu},\\
\Op{HB} &=& ig' (D^\mu H)^\dagger (D^\nu H) B_{\mu\nu},\\
\Op{3W} &=& \frac{g}{6} \epsilon^{abc} W^{a\nu}_\mu W^{b\rho}_\nu W^{c\mu}_\rho.
\eea
We have adopted the notations in Refs.~\cite{Elias-Miro:2013mua,Elias-Miro:2013eta}, and follow the conventions of Peskin and Schroeder~\cite{Peskin:1995ev} where $D_\mu=\partial_\mu-igW^a_\mu\frac{\sigma^a}{2}-ig'B_\mu Y$, $W^a_{\mu\nu} = \partial_\mu W^a_\nu - \partial_\nu W^a_\mu + g\epsilon^{abc}W^b_\mu W^c_\nu$, $W^3_\mu = \cw Z_\mu+\sw A_\mu$, $B_\mu = -\sw Z_\mu+\cw A_\mu$.

While our calculation of $\delb{NP}\Obs{i}{}$ involves only the operators listed above, we emphasize that their Wilson coefficients do not have unambiguous meanings unless the full set of dimension-six operators chosen for the analysis is specified. Usually a complete operator basis is desirable, although in other cases it is helpful to work with an over-complete operator set, as long as one is careful about the RG running of the Wilson coefficients. A complete basis does not have to contain all the operators listed above. For example, the ``EGGM basis'' in Ref.~\cite{Elias-Miro:2013eta} eliminates $\Op{HW}$ and $\Op{HB}$ via integration-by-parts in favor of two operators that affect Higgs physics only:
\bea
\Op{HW} \to \Op{W} -\frac{1}{4} (\Op{WW}+\Op{WB})\quad &\text{where}&\quad \Op{WW} = g^2 |H|^2 W^a_{\mu\nu} W^{a\mu\nu},\\
\Op{HB} \to \Op{B} -\frac{1}{4} (\Op{BB}+\Op{WB})\quad &\text{where}&\quad \Op{BB} = g'^2 |H|^2 B_{\mu\nu} B^{\mu\nu}.
\eea
Other operator bases, such as the one in Ref.~\cite{Grzadkowski:2010es}, trade some of the bosonic operators for those involving fermions via field redefinitions (or equivalently, equations of motion), which makes the definition of universal theories less transparent. We will keep all the operators above in the calculation, so that the final results can be easily adapted to and interpreted in different bases.

\subsection{Calculation of new physics effects}

There are five aspects of the SM calculation of $\eeww$ polarized cross sections $\sigma_{L,R}$ that are affected at LO by the operators listed above.
\begin{itemize}
\item Modifications of $VWW$ vertices ($V=Z,\gamma$). Assuming on-shell $W^+W^-$, the vertex functions, defined by the Feynman rules
\bea
Z_\lambda (q) \to W^+_\mu (k_+) W^-_\nu (k_-) :&\quad& ie\frac{\cw}{\sw} \Gamma_Z^{\mu\nu\lambda} (q,k_+,k_-),\\
A_\lambda (q) \to W^+_\mu (k_+) W^-_\nu (k_-) :&\quad& ie \Gamma_\gamma^{\mu\nu\lambda} (q,k_+,k_-),
\eea
can be parametrized by $q^2$-dependent form factors~\cite{Hagiwara:1986vm}
\beq{VWW}
\Gamma_V^{\mu\nu\lambda} (q,k_+,k_-) = f_1^V(q^2) g^{\mu\nu} (k_--k_+)^\lambda -\frac{f_2^V(q^2)}{v^2} q^\mu q^\nu(k_--k_+)^\lambda +f_3^V(q^2) (q^\nu g^{\lambda\mu} - q^\mu g^{\nu\lambda}).
\eeq
There are additional form factors if C and/or P violation is allowed, but these do not interfere with the LO SM contribution and are thus not considered here. In the EFT, the form factors read
\bea
f_1^V(q^2) &=& 1+\Delta f_1^V(q^2) = 1 + \Delta g_1^V + \frac{q^2}{2v^2} \lambda_V + \frac{e^2}{2\sw^2} \left(\frac{q^2}{2m_W^2}+1\right) c_{2W},\\
f_2^V(q^2) &=& \Delta f_2^V(q^2) = \lambda_V,\\
f_3^V(q^2) &=& 2+\Delta f_3(q^2) = 2 + \Delta g_1^V + \Delta \kappa_V + \frac{m_W^2}{v^2} \lambda_V + \frac{e^2}{\sw^2} \left(\frac{q^2}{2m_W^2}+1\right) c_{2W},
\eea
where
\bea
\Delta g_1^\gamma &=& 0,\\
\Delta g_1^Z &=& -\frac{e^2}{4\cw^2\sw^2} (c_{HW}+c_W),\\
\Delta\kappa_\gamma &=& -\frac{e^2}{4\sw^2} (c_{HW}+c_{HB}-4c_{WB}),\\
\Delta\kappa_Z &=& \Delta g_1^Z -\frac{\sw^2}{\cw^2}\Delta\kappa_\gamma,\\
\lambda_\gamma &=& \lambda_Z = -c_{3W}
\eea
are the commonly-used anomalous TGC parameters. We have rescaled the gauge fields $W^a_\mu$, $B_\mu$ to have their kinetic terms canonically normalized, and simultaneously rescaled $g$, $g'$ so that $gW^a_\mu$, $g'B_\mu$ (and hence the gauge interactions of the fermions) are unchanged\footnote{This is possible at the dimension-six level in the EFT because the kinetic terms for $W^\pm_\mu$ and $W^3_\mu$ are rescaled by the same factor. In other words, dimension-six operators do not generate a nonzero $U$ parameter.}. The weak mixing angle is redefined accordingly to retain $\sw=\frac{g'}{\sqrt{g^2+g'^2}}$, and the mass eigenstate fields are still defined by $Z_\mu = \cw W^3_\mu -\sw B_\mu$, $A_\mu = \sw W^3_\mu +\cw B_\mu$. The term $W^3_{\mu\nu}B^{\mu\nu}$, on the other hand, is not rotated away, which leads to noncanonical normalizations for $Z_\mu$, $A_\mu$, as well as kinetic mixing.
\item Corrections to the $s$-channel $Z/\gamma$ propagators. These can be viewed as corrections to the external leg of the $VWW$ vertices:
\bea
\Gamma_Z^{\mu\nu\lambda} &\to& \left(1+ \frac{q^2}{q^2-m_Z^2}\Delta_{ZZ}(q^2)\right) \Gamma_Z^{\mu\nu\lambda} + \frac{\sw}{\cw} \Delta_{\gamma Z}(q^2) \Gamma_\gamma^{\mu\nu\lambda},\label{ZWWse}\\
\Gamma_\gamma^{\mu\nu\lambda} &\to& \left(1+ \Delta_{\gamma\gamma}(q^2) \right) \Gamma_\gamma^{\mu\nu\lambda} + \frac{\cw}{\sw} \frac{q^2}{q^2-m_Z^2}\Delta_{\gamma Z}(q^2) \Gamma_Z^{\mu\nu\lambda},\label{AWWse}
\eea
where
\bea
\Delta_{ZZ} (q^2) &\equiv& \frac{\Pi_{ZZ}(q^2)}{q^2} = -\frac{m_Z^2}{q^2} c_T + \frac{e^2}{2} (4c_{WB}+c_W+c_B) -\frac{q^2}{v^2} (\cw^2 c_{2W} + \sw^2 c_{2B}),\\
\Delta_{\gamma Z} (q^2) &\equiv& \frac{\Pi_{\gamma Z}(q^2)}{q^2} = - \frac{\cw^2-\sw^2}{4\cw\sw}e^2 (4c_{WB}+c_W+c_B) -\frac{q^2}{v^2}\cw\sw (c_{2W} - c_{2B}),\\
\Delta_{\gamma\gamma} (q^2) &\equiv& \frac{\Pi_{\gamma\gamma}(q^2)}{q^2} = -\frac{e^2}{2} (4c_{WB}+c_W+c_B) -\frac{q^2}{v^2} (\sw^2 c_{2W} + \cw^2 c_{2B})
\eea
are self-energy corrections due to new physics. Here and in the following, $\Pi_{VV'}$ represent the new physics contribution to the self-energies. Since in the SM, $\Gamma_Z^{\mu\nu\lambda}=\Gamma_\gamma^{\mu\nu\lambda}$, \eqs{ZWWse}{AWWse} are equivalent to an additional contribution to the form factors:
\bea
\Delta_\text{se} f_1^Z (q^2) &=& \frac{q^2}{q^2-m_Z^2} \Delta_{ZZ}(q^2) + \frac{\sw}{\cw} \Delta_{\gamma Z}(q^2),\\
\Delta_\text{se} f_1^\gamma (q^2) &=& \Delta_{\gamma\gamma}(q^2) + \frac{\cw}{\sw} \frac{q^2}{q^2-m_Z^2} \Delta_{\gamma Z}(q^2),\\
\Delta_\text{se} f_2^V (q^2) &=& 0,\\
\Delta_\text{se} f_3^V (q^2) &=& 2\Delta_\text{se} f_1^V (q^2),
\eea
where the subscript ``se'' stands for ``self-energy''.
\item Shifts in $m_W$ which enters the kinematics. With the Higgs VEV rescaled to leave the $W$ boson mass term $\left(\frac{gv}{2}\right)^2 W^+_\mu W^{-\mu}$ unchanged, only $\Op{2W}$ contributes to $m_W$ via self-energy corrections
\be
m_W^2 \to m_W^2 + \Pi_{WW}(m_W^2) = m_W^2 \left(1 - \frac{e^2}{4\sw^2} c_{2W}\right),
\ee
where $m_W$ should be understood as a shorthand for $\frac{gv}{2}$. We emphasize that this step is essential regardless of whether $m_W$ is in the input observables set because the direct new physics corrections $\xi_{\sigma_{L,R}}$ are defined with respect to the Lagrangian parameters (there are cancellations between direct and indirect contributions if $m_W$ is an input observable). There are also shifts in $m_Z$, but these are already contained in the propagator corrections calculated above.
\item The $W^\pm$ field strength renormalization factors. The cross sections are simply rescaled.
\be
\sigma_{L,R} \to \sigma_{L,R} \left(1+2\Pi_{WW}'(m_W^2)\right) = \sigma_{L,R} \left(1-\frac{e^2}{\sw^2} c_{2W}\right).
\ee
\item Indirect contributions via shifts in the input observables. Using the input observables set $\{m_Z, G_F, \alpha\}$, we have
\bea
\xi_{m_Z} &=& -\frac{1}{2} c_T + \frac{e^2}{4} (4c_{WB}+c_W+c_B) -\frac{e^2}{8\cw^2\sw^2} (\cw^2 c_{2W} + \sw^2 c_{2B}), \\
\xi_\alpha &=& -\frac{e^2}{2} (4c_{WB}+c_W+c_B).
\eea
$\xi_{G_F} = 0$ because $v$ has been rescaled such that $\Pi_{WW}(0)=0$.
\end{itemize}

Assembling all the pieces, we arrive at the final result
\bea
\delb{NP}\sigma_{L,R} = \frac{1}{\sigma_{L,R}} \left[ \sum_{i,V} \frac{\partial\sigma_{L,R}}{\partial f_i^V} (\Delta f_i^V + \Delta_\text{se} f_i^V) +\frac{\partial\sigma_{L,R}}{\partial m_W^2}\Pi_{WW}(m_W^2) \right]&& \nonumber\\
+2\Pi_{WW}'(m_W^2) -d_{\sigma_{L,R},m_Z}\xi_{m_Z} -d_{\sigma_{L,R},\alpha}\xi_{\alpha}.&&\label{assem}
\eea
Explicit expressions for $\sigma_{L,R}$ and $\frac{\partial\sigma_{L,R}}{\partial f_i^V}$ can be found in Appendix~\ref{app:LO}. Up to $\Ord{\frac{v^2}{\Lambda^2}\alpha}$ corrections, we can trade the parameters in the final result of the calculation for the input observables $\{m_Z, G_F, \alpha\}$ through LO relations, and use the LO results for $d_{\sigma_{L,R},i'}$ (see Fig.~\ref{fig:NLOvsLO} for the size of NLO corrections to $d_{\sigma_{L,R},i'}$). Numerically, using the reference values listed in Section~\ref{sec:param} for the input observables, we obtain, for $\sqrt{s}=200~\text{GeV}$,
\bea
\delb{NP}\sigma_L &=& 0.0192 (c_{HW}+c_W) +0.00345 (c_{HW}+c_{HB}-4c_{WB}) +0.00667 c_{3W} \nonumber\\
&& -0.0967 (4c_{WB}+c_W+c_B) +1.66 c_T -0.183 c_{2W} +0.0442 c_{2B},\label{sigL200}\\
\delb{NP}\sigma_R &=& -1.32(c_{HW}+c_W) +0.640 (c_{HW}+c_{HB}-4c_{WB}) -0.0898 c_{3W} \nonumber\\
&& +1.67 (4c_{WB}+c_W+c_B) -5.45 c_T -0.173 c_{2W} -1.49 c_{2B},\\
\delb{NP}\sigma &=& 0.00513 (c_{HW}+c_W) +0.0101 (c_{HW}+c_{HB}-4c_{WB}) +0.00566 c_{3W} \nonumber\\
&& -0.0782 (4c_{WB}+c_W+c_B) +1.58 c_T -0.183 c_{2W} +0.0281 c_{2B},
\eea
and for $\sqrt{s}=500~\text{GeV}$,
\bea
\delb{NP}\sigma_L &=& 0.0835 (c_{HW}+c_W) +0.0277 (c_{HW}+c_{HB}-4c_{WB}) +0.0191 c_{3W} \nonumber\\
&& -0.115 (4c_{WB}+c_W+c_B) +2.38 c_T -0.253 c_{2W} +0.0396 c_{2B},\\
\delb{NP}\sigma_R &=& -8.25 (c_{HW}+c_W) +6.64 (c_{HW}+c_{HB}-4c_{WB}) -0.0426 c_{3W} \nonumber\\
&& +8.41 (4c_{WB}+c_W+c_B) -2.61 c_T -0.0826 c_{2W} -8.33 c_{2B},\\
\delb{NP}\sigma &=& 0.0497 (c_{HW}+c_W) +0.0546 (c_{HW}+c_{HB}-4c_{WB}) +0.0189 c_{3W} \nonumber\\
&& -0.0804 (4c_{WB}+c_W+c_B) +2.36 c_T -0.252 c_{2W} +0.00563 c_{2B}.\label{sig500}
\eea
We have also shown the new physics contributions to the unpolarized cross sections in the equations above, which are directly calculated from
\be
\delb{NP}\sigma = \frac{\sigma_L}{\sigma_L+\sigma_R} \delb{NP}\sigma_L + \frac{\sigma_R}{\sigma_L+\sigma_R} \delb{NP}\sigma_R.
\ee
The results for the left-right asymmetries are derived from
\bea
\delb{NP}\AbLR &=& \delb{NP}\sigma_R - \delb{NP}\sigma,\\
\delb{NP}\ALR &=& -\frac{\AbLR}{1-\AbLR}\delb{NP}\AbLR =
\begin{cases}
-0.0214~\delb{NP}\AbLR \,\,\text{for}\,\,\sqrt{s}=200~\text{GeV}\\
-0.00818~\delb{NP}\AbLR \,\,\text{for}\,\,\sqrt{s}=500~\text{GeV}
\end{cases}
\,\,\text{(LO)},
\eea
and will not be listed explicitly.

\subsection{Interpretation of results}

To interpret these results, namely to see the role played by $\eeww$ in probing the EFT parameter space, we need to go to a specific basis, and compare our results with other experimental constraints on the Wilson coefficients {\it in this basis}. We consider the EGGM basis~\cite{Elias-Miro:2013eta} for illustration, and relate four combinations of Wilson coefficients in this basis to the oblique parameters $\hat S, \hat T, W, Y$~\cite{Peskin:1991sw,Maksymyk:1993zm,Barbieri:2004qk}:
\bea
\hat S &\equiv& \frac{\alpha S}{4\sw^2} = -\frac{\cw}{\sw} \Pi_{3B}'(0) = \frac{e^2}{4\sw^2} (4c_{WB}+c_W+c_B),\\
\hat T &\equiv& \alpha T = \frac{1}{m_W^2} \bigl[\Pi_{WW}(0)-\Pi_{33}(0)\bigr] = c_T,\\
W &\equiv& -\frac{m_W^2}{2} \Pi_{33}''(0) = \frac{e^2}{4\sw^2} c_{2W},\\
Y &\equiv& -\frac{m_W^2}{2} \Pi_{BB}''(0) = \frac{e^2}{4\sw^2} c_{2B}.
\eea
We have adopted the definitions in Ref.~\cite{Henning:2014wua}, which differ from Ref.~\cite{Barbieri:2004qk} by sign. In universal theories, these four parameters are sufficient to describe the $Z$-pole data at LEP1, $m_W$ measurements, and $e^+e^-\to f\bar f$ data at LEP 2~\cite{Barbieri:2004qk}. Each of them is constrained at the $10^{-3}$ level~\cite{Barbieri:2004qk}. It is clear that these oblique parameters are not sensitive to $c_{3W}$. Also, they constrain only one combination of the three coefficients $c_{WB}, c_W, c_B$. Choosing the other two combinations to be $c_{WB}$ and $c_W-c_B$, we rewrite Eqs.~(\ref{sigL200}-\ref{sig500}) in the EGGM basis as follows.
\bea
\delb{NP}\sigma_L (\sqrt{s}=200~\text{GeV}) &=& -0.0521 c_{WB} +0.00958 (c_W-c_B) +0.00667 c_{3W} \nonumber\\
&& -0.826 \hat S +1.66 \hat T -1.74 W +0.419 Y,\label{sigL200EGGM}\\
\delb{NP}\sigma_R (\sqrt{s}=200~\text{GeV}) &=& 0.0805 c_{WB} -0.660 (c_W-c_B) -0.0898 c_{3W} \nonumber\\
&& +9.53 \hat S -5.45 \hat T -1.64 W -14.1 Y,\\
\delb{NP}\sigma (\sqrt{s}=200~\text{GeV}) &=& -0.0507 c_{WB} +0.00256 (c_W-c_B) +0.00566 c_{3W} \nonumber\\
&& -0.717 \hat S +1.58 \hat T -1.74 W +0.267 Y,\\
\delb{NP}\sigma_L (\sqrt{s}=500~\text{GeV}) &=& -0.278 c_{WB} +0.0418 (c_W-c_B) +0.0191 c_{3W} \nonumber\\
&& -0.695 \hat S +2.38 \hat T -2.40 W +0.375 Y,\\
\delb{NP}\sigma_R (\sqrt{s}=500~\text{GeV}) &=& -10.1c_{WB} -4.12 (c_W-c_B) -0.0426 c_{3W} \nonumber\\
&& +40.7 \hat S -2.61 \hat T -0.783 W -79.0 Y,\\
\delb{NP}\sigma (\sqrt{s}=500~\text{GeV}) &=& -0.318 c_{WB} +0.0249 (c_W-c_B) +0.0189 c_{3W} \nonumber\\
&& -0.527 \hat S +2.36 \hat T -2.39 W +0.0534 Y.\label{sig500EGGM}
\eea
It should be noted that the coefficients of $\hat S$ in these equations are not unique, as they depend on the choice for the other two combinations among $c_{WB}, c_W, c_B$. Our choice is motivated by the observation that in weakly-coupled new physics scenarios, $\Op{W}$ and $\Op{B}$ are ``potential-tree-generated'', while $\Op{WB}$ is ``loop-generated''~\cite{Arzt:1994gp,Giudice:2007fh,Elias-Miro:2013gya,Elias-Miro:2013mua,Einhorn:2013kja,Einhorn:2013tja}. In the limit $|c_{WB}|\ll |c_W|, |c_B|$, the $\hat S$ parameter involves $c_W+c_B$, while the orthogonal combination $c_W-c_B$ can be probed by $\eeww$.

Eqs.~(\ref{sigL200EGGM}-\ref{sig500EGGM}) demonstrate the complementarity of $\eeww$ and other precision data in probing the parameter space of universal theories at the observables level. As such they are free from ambiguities and additional assumptions. It is seen that cross sections at $\sqrt{s}=500$~GeV tend to have stronger dependence on the Wilson coefficients, and hence better sensitivity to new physics effects. This is not surprising since there are contributions that scale as $\frac{s}{\Lambda^2}$, a fact that is easier to see in a basis where operators involving fermions are retained~\cite{Buchalla:2013wpa}. We have chosen to focus on bases that maximize the use of bosonic operators for easier connection with the literature on universal theories, but one should not misinterpret the relatively strong dependence on multiple Wilson coefficients in Eqs.~(\ref{sigL200EGGM}-\ref{sig500EGGM}) as necessarily indicating experimental sensitivity to all of them. To draw quantitative conclusions about the constraining power of $\eeww$, a global analysis should be carried out, where correlations among the constraints on different Wilson coefficients become clear. The calculations and results presented in this paper can be used as a starting point.

\section{Conclusions}\label{sec:conclusions}

Precision electroweak analyses will continue to contribute to our understanding of Nature in the post-Higgs-boson-discovery era. A truly consistent approach to precision analyses, both as a consistency test of the SM and as an indirect probe of new physics, requires not only the experimental and theoretical precision, but also the dependence of the calculation on the input observables to be well understood. Though in many cases parametric dependence is insignificant compared with experimental and theoretical uncertainties, it should not be taken for granted that it will remain so, especially when the projected experimental precisions greatly exceed the current ones. Instead, careful justification is needed.

In this article, we have investigated the role of $\eeww$ in the precision program, motivated by the projected per-mil-level cross section measurements and the complementarity between this and other processes in probing new physics effects. The latter is also relevant for precision studies of the Higgs boson. We have utilized the best SM calculations available as public codes, and presented the results for several inclusive observables in terms of expansion formulas. From these one can directly read off the dependence on the input observables, which allows us to justify the exclusion of inclusive $\eeww$ observables from the global precision electroweak analysis as a test of the SM. In fact, even in the future, the experimental uncertainties are still expected to dominate over the parametric uncertainties, indicating an insignificant contribution to the $\chi^2$ function. Our analysis also justifies the neglect of SM parametric uncertainties when using precision data to constrain new physics contributions above the $10^{-4}$ level.

On the other hand, to demonstrate the interplay between $\eeww$ observables and other precision data in probing new physics effects, we have considered a representative class of models, the universal theories. These models have been extensively studied in the oblique parameters framework, where four parameters suffice to incorporate the most stringent experimental constraints. For $\eeww$, we have gone beyond the TGC parametrization, and calculated well-defined physical observables in the EFT framework, again using the expansion formalism. While some terms in the expansion can be mapped to the oblique parameters, other terms show the additional directions in the EFT parameter space accessible to $\eeww$. Our calculations and results provide necessary tools for a consistent global analysis, which will tell us where we are in the EFT parameter space if deviations from the SM are found in upcoming precision measurements, and will thus give us guidance for where to look for new physics if it fails to show up directly.

\acknowledgments
The authors acknowledge support in part by the DoE under grants DE-SC0007859 and DE-SC0011719.

\appendix

\section{Leading-order results for $e^+e^-\to W^+W^-$ cross sections in the SM}\label{app:LO}
At LO, the full $e^+e^-\to 4f$ process factorizes into $W^+W^-$ production and decay. The decay branching fractions of $W^+$ (and same for $W^-$) simply follow from final state counting, and are $1/3$ for hadronic final states and $1/9$ for leptonic final states. In the PDG notation~\cite{Agashe:2014kda}, the production cross sections for polarized $e^-$ read
\bea
\sigma_L (s; m_Z^2, m_W^2, e^2, \sw^2) &=& \frac{e^4}{8\pi s^2} \int_{t_+}^{t_-} \diff t \Biggl[ \left(\frac{2\sw^2-1}{2\sw^2}\frac{s}{s-m_Z^2}-1\right)^2 A(s,t;m_W^2) \nonumber\\
&&+ \frac{1}{\sw^2}\left(\frac{2\sw^2-1}{2\sw^2}\frac{s}{s-m_Z^2}-1\right) I(s,t;m_W^2) + \frac{1}{4\sw^4} E(s,t;m_W^2)\Biggr], \label{sigLSM}\\
\sigma_R (s; m_Z^2, m_W^2, e^2) &=& \frac{e^4}{8\pi s^2} \int_{t_+}^{t_-} \diff t \left(\frac{s}{s-m_Z^2}-1\right)^2 A(s,t;m_W^2),\label{sigRSM}
\eea
where $t_\pm = m_W^2-\frac{s}{2}\left(1-\sqrt{1\pm\frac{4m_W^2}{s}}\right)$. 
The functions $A, I, E$ come from the $s$-channel, interference, and $t$-channel contributions, respectively. Explicitly,
\bea
A(s,t;m_W^2) &=& -\frac{s^2}{4m_W^4} \left(\frac{t}{s}+\frac{t^2}{s^2}\right) +\frac{s}{2m_W^2} \left(2+\frac{3t}{s}+\frac{2t^2}{s^2}\right) -\frac{1}{4} \left(17+\frac{20t}{s}+\frac{12t^2}{s^2}\right) \nonumber\\
&& +\frac{m_W^2}{s} \left(1+\frac{6t}{s}\right) -\frac{3m_W^4}{s^2}, \\
I(s,t;m_W^2) &=& -\frac{s^2}{4m_W^4} \left(\frac{t}{s}+\frac{t^2}{s^2}\right) +\frac{s}{2m_W^2} \left(2+\frac{2t}{s}+\frac{t^2}{s^2}\right) -\frac{5}{4} \nonumber\\
&& +\frac{m_W^2}{2s} \left(\frac{4s}{t}-3\right) +\frac{m_W^4}{st}, \\
E(s,t;m_W^2) &=& -\frac{s^2}{4m_W^4} \left(\frac{t}{s}+\frac{t^2}{s^2}\right) +\frac{s}{2m_W^2} \left(2+\frac{t}{s}\right) 
 -\frac{1}{4} \left(\frac{4s}{t}+5\right) +\frac{2m_W^2}{t} -\frac{m_W^4}{t^2}.
\eea

\eqs{sigLSM}{sigRSM} are written in terms of the parameters that directly enter the calculation, all of which are not independent. This parametrization is convenient for the calculation of direct contributions from new physics, namely $\xi_{\sigma_{L,R}}$. But regarding the indirect contributions $-\sum_{i'} d_{\sigma_{L,R}, i'} \xi_{i'}$, one should be careful not to over-count the number of independent inputs. Using $\{m_Z, G_F, \alpha \}$ as input observables, we have
\bea
d_{\sigma_L,m_Z} &=& \frac{2m_Z^2}{\sigma_L} \left(\frac{\partial\sigma_L}{\partial m_Z^2} + \frac{\partial\sigma_L}{\partial m_W^2}\frac{\partial m_W^2}{\partial m_Z^2} + \frac{\partial\sigma_L}{\partial\sw^2}\frac{\partial\sw^2}{\partial m_Z^2}\right),\\
d_{\sigma_L,G_F} &=& \frac{G_F}{\sigma_L} \left(\frac{\partial\sigma_L}{\partial m_W^2}\frac{\partial m_W^2}{\partial G_F} + \frac{\partial\sigma_L}{\partial\sw^2}\frac{\partial\sw^2}{\partial G_F}\right),\\
d_{\sigma_L,\alpha} &=& 2+\frac{\alpha}{\sigma_L} \left(\frac{\partial\sigma_L}{\partial m_W^2}\frac{\partial m_W^2}{\partial \alpha} + \frac{\partial\sigma_L}{\partial\sw^2}\frac{\partial\sw^2}{\partial \alpha}\right),
\eea
where
\bea
m_W^2(m_Z^2, G_F, \alpha) &=& \frac{m_Z^2}{2} \left(1+ \sqrt{1-\frac{4\pi\alpha}{\sqrt{2}G_F m_Z^2}}\right),\\
\sw^2(m_Z^2, G_F, \alpha) &=& \frac{1}{2} \left(1- \sqrt{1-\frac{4\pi\alpha}{\sqrt{2}G_F m_Z^2}}\right)
\eea
are used to calculate the partial derivatives involved. For $\sigma_R$, just drop the terms with $\sw^2$. It is understood that any dependence on the calculational inputs in the final results should be traded for the three input observables, which can be done at LO, before numerical values for the input observables are plugged in. These LO results for the expansion coefficients are also used for comparison with the results from \rac\ in Fig.~\ref{fig:NLOvsLO}.

It is shown in Sec.~\ref{sec:universal} that a large part of new physics effects is effectively characterized by the form factors defined in \eq{VWW}. The partial derivatives $\frac{\partial\sigma_{L,R}}{\partial f_i^V}$ in \eq{assem} read
\bea
\frac{\partial \sigma_L}{\partial f_i^Z} &=& \frac{e^4}{8\pi s^2} \int_{t_+}^{t_-} \diff t \Biggl[ \left(\frac{2\sw^2-1}{2\sw^2}\frac{s}{s-m_Z^2}-1\right) \left(\frac{2\sw^2-1}{2\sw^2}\frac{s}{s-m_Z^2}\right) \frac{\partial A}{\partial f_i} \nonumber\\
&&\qquad\qquad\qquad + \frac{1}{\sw^2}\left(\frac{2\sw^2-1}{2\sw^2}\frac{s}{s-m_Z^2}\right) \frac{\partial I}{\partial f_i} \Biggr], \\
\frac{\partial \sigma_L}{\partial f_i^\gamma} &=& \frac{e^4}{8\pi s^2} \int_{t_+}^{t_-} \diff t \Biggl[ \left(\frac{2\sw^2-1}{2\sw^2}\frac{s}{s-m_Z^2}-1\right) (-1) \frac{\partial A}{\partial f_i} + \frac{1}{\sw^2}(-1) \frac{\partial I}{\partial f_i} \Biggr], \\
\frac{\partial \sigma_R}{\partial f_i^Z} &=& \frac{e^4}{8\pi s^2} \int_{t_+}^{t_-} \diff t \left(\frac{s}{s-m_Z^2}-1\right) \left(\frac{s}{s-m_Z^2}\right) \frac{\partial A}{\partial f_i}, \\
\frac{\partial \sigma_R}{\partial f_i^\gamma} &=& \frac{e^4}{8\pi s^2} \int_{t_+}^{t_-} \diff t \left(\frac{s}{s-m_Z^2}-1\right) (-1) \frac{\partial A}{\partial f_i},
\eea
where $\frac{\partial A}{\partial f_i}$, $\frac{\partial I}{\partial f_i}$ are obtained by setting $f_i^Z=f_i^\gamma=f_i$ and taking derivatives,
\bea
\frac{\partial A}{\partial f_1} &=& \frac{s^2}{2m_W^4} \left(\frac{t}{s}+\frac{t^2}{s^2}\right) -\frac{t}{m_W^2} +\frac{1}{2}\left(1-\frac{12t}{s}-\frac{12t^2}{s^2}\right) +\frac{12m_W^2 t}{s^2} -\frac{6m_W^4}{s^2},\\
\frac{\partial A}{\partial f_2} &=& \frac{s}{v^2} \Biggl[ -\frac{s^2}{4m_W^4} \left(\frac{t}{s}+\frac{t^2}{s^2}\right) +\frac{s}{2m_W^2} \left(\frac{2t}{s}+\frac{t^2}{s^2}\right) -\frac{1}{4}\left(1-\frac{4t}{s}-\frac{8t^2}{s^2}\right) \nonumber\\
&&+\frac{m_W^2}{2s} \left(1-\frac{8t}{s}\right) +\frac{2m_W^4}{s^2} \Biggr],\\
\frac{\partial A}{\partial f_3} &=& -\frac{s^2}{2m_W^4} \left(\frac{t}{s}+\frac{t^2}{s^2}\right) +\frac{s}{m_W^2} \left(1+\frac{2t}{s}+\frac{t^2}{s^2}\right) -\frac{1}{2} \left(9+\frac{4t}{s}\right) +\frac{m_W^2}{s}, \\
\frac{\partial I}{\partial f_1} &=& \frac{s^2}{4m_W^4} \left(\frac{t}{s}+\frac{t^2}{s^2}\right) +\frac{t^2}{2m_W^2 s} +\frac{5}{4} -\frac{3m_W^2}{2s} +\frac{m_W^4}{st}, \\
\frac{\partial I}{\partial f_2} &=& \frac{s}{v^2} \Biggl[ -\frac{s^2}{8m_W^4} \left(\frac{t}{s}+\frac{t^2}{s^2}\right) +\frac{t}{4m_W^2} -\frac{1}{2}\left(\frac{5}{4}+\frac{t}{s}\right) +\frac{m_W^2}{s} -\frac{m_W^4}{2st} \Biggr], \\
\frac{\partial I}{\partial f_3} &=& -\frac{s^2}{4m_W^4} \left(\frac{t}{s}+\frac{t^2}{s^2}\right) +\frac{s}{2m_W^2}\left(1+\frac{t}{s}\right) -\frac{5}{4} +\frac{m_W^2}{t}.
\eea

\section{Technical details of parametric dependence calculations}\label{sec:tech}

Among the five observables introduced in Sec.~\ref{sec:obs}, only two are independent, which we choose to be $\sigma$ and $\sigma_R$ for the present calculation. The reference values and expansion coefficients for the other three observables can be derived from those for $\sigma$, $\sigma_R$ as follows:
\begin{align}
\sigma_L^{\text{ref}} &= 2\sigma^{\text{ref}} -\sigma_R^{\text{ref}},& c_{\sigma_L,i'} &= \frac{1}{\sigma_L^{\text{ref}}} (2\sigma^{\text{ref}} c_{\sigma,i'} - \sigma_R^{\text{ref}} c_{\sigma_R,i'}),\\
\ALR^{\text{ref}} &= 1- \frac{\sigma_R^{\text{ref}}}{\sigma^{\text{ref}}},& c_{\ALR,i'} &= \left(\frac{1}{\ALR^{\text{ref}}}-1\right) (c_{\sigma,i'} - c_{\sigma_R,i'}), \\
\AbLR^{\text{ref}} &= \frac{\sigma_R^{\text{ref}}}{\sigma^{\text{ref}}},& c_{\AbLR,i'} &= c_{\sigma_R,i'} - c_{\sigma,i'}.
\end{align}

We use the \rac\ package to calculate $e^+e^-\to u\bar d \mu^-\bar\nu_\mu$ cross sections $\sigma$, $\sigma_R$, adopting the recommended choices for the switches in the input file but removing all separation cuts. The program is run multiple times with shifted input observables before the expansion coefficients $c_{i,i'}$ are calculated by the finite difference method,
\beq{cfd}
c_{i,i'} = \frac{\Obs{i'}{ref}}{\Obs{i}{ref}} \left.\frac{\partial \Obs{i}{SM}}{\partial \Obs{i'}{}}\right|_{\Obs{i'}{}=\Obs{i'}{ref}}
= \frac{\left.\Obs{i}{SM}\right|_{(1+h_{i,i'})\Obs{i'}{ref}} - \left.\Obs{i}{SM}\right|_{(1-h_{i,i'})\Obs{i'}{ref}}}{2h_{i,i'}\Obs{i}{ref}} + \Ord{h^2}.
\eeq
This is clearly CPU-time-consuming since the calculation involves Monte Carlo integration over the final-state four-fermion phase space. We thus keep the number of weighted events generated per run of the program $N_{\text{events}}$ at a minimum that still allows interesting information to be extracted. In particular, for the extraction of $c_{i,m_Z}$, $c_{i,G_F}$, $c_{i,m_W}$, we combine the results from the two branches ``slicing'' and ``subtraction'' of \rac, with $N_{\text{events}}=10^7$ for each branch, which is the recommended minimum. The two branches differ in the treatment of IR singularities, and they are found to yield compatible results. For the much smaller $c_{i,m_t}$, $c_{i,\alpha_s}$, $c_{i,m_H}$, we set $N_{\text{events}}=5\times10^7$ to further reduce Monte Carlo error, while running the faster ``slicing'' branch only. The reference values $\sigma^{\text{ref}}$, $\sigma_R^{\text{ref}}$ are also obtained with the ``slicing'' branch with $N_{\text{events}}=5\times10^7$.

The choices of $h_{i,i'}$ in \eq{cfd} are listed in Table~\ref{table:h}. These $h$'s are small enough so that reducing $h$ further leads to larger Monte Carlo error bars that encompass the current ones. In other words, the truncation errors due to the discarded $\Ord{h^2}$ term are negligible compared with the Monte Carlo errors. Of course, the optimal $h$'s would be smaller if the Monte Carlo errors are reduced by increasing $N_{\text{events}}$. The Monte Carlo errors quoted in Tables~\ref{table:c} and~\ref{table:d} are obtained assuming individual runs of \rac\ yield uncorrelated errors. This assumption is probably very conservative because, since the same random number seed is used for each run, the results for $\sigma$, $\sigma_R$ for slightly different values of the input observables tend to fluctuate in a correlated way. Therefore, the central values of the expansion coefficients shown in Tables~\ref{table:c} and~\ref{table:d} are likely to be more robust than the error bars suggest.
\begin{table}
\begin{center}
  \begin{tabular}{|c|cccccc|}
    \hline
     \hspace{10pt}$\Obs{i}{}$\hspace{10pt} & \hspace{10pt}$h_{i,m_Z}$\hspace{10pt} & \hspace{10pt}$h_{i,G_F}$\hspace{10pt} & \hspace{10pt}$h_{i,m_W}$\hspace{10pt} & \hspace{10pt}$h_{i,m_t}$\hspace{10pt} & \hspace{10pt}$h_{i,\alpha_s}$\hspace{10pt} & \hspace{10pt}$h_{i,m_H}$\hspace{10pt} \\
    \hline
    $\sigma$ & 0.03 & 0.03 & 0.03 & 0.1 & 0.1 & 0.1 \\
    \hline
    $\sigma_R$ & 0.005 & 0.03 & 0.005 & 0.1 & 0.1 & 0.05/0.1 \\
    \hline
  \end{tabular}
\end{center}
\caption{The choices for $h_{i,i'}$ in \eq{cfd}, the fractional shift in the input $\Obs{i'}{}$ with respect to $\Obs{i'}{ref}$ for the extraction of $c_{i,i'}$. For $h_{\sigma_R, m_H}$, 0.05 is used for $\sqrt{s}=200$~GeV while 0.1 is used for $\sqrt{s}=500$~GeV.}\label{table:h}
\end{table}

\end{document}